# STREAMER SELF-FOCUSING
# IN EXTERNAL LONGITUDINAL MAGNETIC FIELD

A Yu Starikovskiy[1,*] N L Aleksandrov[2], M N Shneider[1]

[1]Princeton University, Princeton, NJ08544, USA
[2]Moscow Institute of Physics and Technology, Dolgoprudny, 141700, Russia
[*]Author to whom any correspondence should be addressed
E-mail: astariko@princeton.edu

**Abstract**

The numerical simulation of the development of a streamer discharge in a gap with an external longitudinal magnetic field was used to demonstrate the self-focusing of such discharges. Self-focusing is caused by a sharp deceleration of the radial ionization wave due to a change in the electron energy distribution function, a decrease in the average electron energy, the rate of gas ionization and the electron mobility in crossed electric and magnetic fields as compared to the case of the discharge development without a magnetic field. The self-focusing effect of a streamer discharge in an external longitudinal magnetic field is observed for both positive and negative pulse polarities. The paper proposes an estimate of the critical value of the magnetic field, which makes it possible to control the development of pulsed high-voltage discharges at various gas pressures.

**Keywords**

nanosecond discharge, streamer, magnetic field, electric field, electron energy distribution function, electron mobility, ionization coefficient, nonequilibrium plasma, Hall parameter, reduced gyrofrequency

**Contents**

**Nomenclature**

$B$ = magnetic flux density
$E$ = electric field
$\alpha$ = angle between the vectors $\boldsymbol{B}$ and $\boldsymbol{E}$
$e$ = elementary charge
$I_D$ = discharge current
$m_e$ = electron mass
$n_e$ = electron number density
$n_i$ = ion number density
$T_e$ = effective electron temperature
$u(\varepsilon)$ = electron velocity
$u_e$ = electron drift velocity
$\varepsilon$ = electron energy
$\langle\varepsilon\rangle$ = average electron energy
$\varphi$ = plasma potential
$\mu_e$ = electron mobility
$\nu_e$ = total electron momentum transfer frequency
$\nu_m(\varepsilon)$ = momentum transfer frequency for electrons with energy $\varepsilon$
$Q_m(\varepsilon)$ = electron momentum cross section
$\beta_e$ = electron Hall parameter
$\omega_e$ = electron gyrotron frequency
$S_{ion}$ = ionization rate
$S_{photo}$ = rate of photoionization
$S_{att}$ = rate of attachment
$S_{rec}^{ei}$ = rate of electron-ion recombination
$S_{rec}^{ii}$ = rate of ion-ion recombination

**Subscripts**

$R$ = radial direction
$Z$ = axial direction
$\perp$ = cross-field direction
$\parallel$ = parallel-field direction

**Introduction**

Magneto-hydrodynamic (MHD) generators might be able to extract significant levels of power from the flow [1] to enable a direct electric power generation from a high-speed flow of combustion products and new high-power demand technologies including plasma steering, plasma-assisted drag reduction, combustion control, and suppression of shock induced separation, all without the need for moving parts. The viability of MHD flow control and power extraction is expected to improve with increasing the flight altitude and velocity. The need to ionize a flow to make it conductive is perhaps the most significant challenge associated with MHD devices operating in Mach number regimes below about Mach 12 [2–4]. In this regime, even the viscous portions of the flow are cold enough such that seeding the flow with an alkali metal vapor will not lead to significant conductivity [2−4]. It has been shown that nonequilibrium ionization methods including electron beams or short, high-voltage pulses are the most efficient means of generating conductivity through the electron-impact gas ionization [2−6]. It has also been shown that by using a nonequilibrium ionization for MHD power extraction the amount of power that is coupled out of a hypersonic flow can be significantly higher than the theoretical power requirements for ionization [2−4]. Because of the large density gradients and resultant conductivity gradients associated with supersonic flows, volume-filling supersonic discharges are often difficult to produce in a wind tunnel.

In addition to MHD flow control and power generation, there are other applications that render such a discharge desirable. Carbon monoxide and excimer gas discharge lasers in general can benefit greatly from the use of a high throughput of cooled (in a supersonic nozzle) gas. To this end, supersonic discharges for pumping such lasers have been reported in molecular flows using DC discharges [7−9], electron beam stabilization techniques [10], and a 13.56-MHz RF discharge [11]. Recent progress has been made at the Ohio State University also using a 13.56-MHz RF source to drive a volume-filling discharge in Mach 2.5 airflow [12]. Pulsed ionization schemes in conjunction with DC discharges have also been implemented successfully in subsonic flows to reduce arcing in $CO_2$ lasers. This has been demonstrated experimentally by Generalov et al. [13−15].

Electron-neutral collisions and their effect on the electrical conductivity strongly influence the nature of MHD interaction with the flow. The two most relevant parameters in assessing the effects of collisions in a nonthermally ionized MHD generator are the electron Hall parameter $\beta_e$ and the electron loss rate. The Hall parameter is defined as the ratio of the electron gyrotron frequency $\omega_e = eB/m_e$ to the electron transport collision frequency $\nu_e$ [16]. In air, the electron loss rate is dominated by attachment to oxygen at low temperatures and high gas densities and by dissociative electron-ion recombination at high ionization levels and lower gas densities [2−5]. The electron loss rate controls the energy required for nonthermal gas ionization.

Application of nanosecond high-voltage pulses to maintain uniform volumetric ionization in a supersonic flow is promising for use in MHD generators. Sustaining conductivity by high-voltage pulses has the great advantage of removing the requirement of the flow seeding and potentially extending the operability of MHD to lower temperatures and Mach numbers, since the conductivity is no longer directly related to thermal ionization of a low ionization potential material such as potassium. High voltage pulses applied along the magnetic field lines play a dual role. First, a sufficiently high degree of ionization and uniformity of the generated plasma in the gas flow are maintained, and, second, the recombination decay of the plasma is significantly slowed down due to the periodic heating of electrons by the electric field of a high-voltage pulsed discharge. Such a scheme of an MHD generator supported by high-voltage pulses in a cold air flow was first implemented in 2006 at Princeton [17]. In this work, MHD power extraction using nonequilibrium ionization in a cold supersonic airflow has been observed. A volumetric, uniform, nonequilibrium cold plasma has been produced in a Mach 3 airflow using 2 ns, 100 kHz repetition rate, 30 kV pulses. Theoretical analysis indicated the electron number density to be on the order of $5\times10^{11}$–$10^{12}$ cm$^{-3}$. The 5 T magnetic field was shown to improve the uniformity of the plasma and had a dramatic effect in confining the plasma to the interelectrode volume.

Thus, the development of nanosecond pulsed discharges in a strong external magnetic field is of significant interest both from the point of view of energy generation in the MHD cycle (for example, when burning natural coal in oxygen) and for the problems of possible $CO_2$ utilization using alternative renewable energy sources. However, for the successful operation of the MHD generator, the high efficiency of the homogeneous plasma production in the flow is required. When ionizing high-voltage pulses are applied, inhomogeneities can appear and develop during the pulse, creating thin plasma channels with a relatively high degree of ionization in the boundary layers and in the flow core. These inhomogeneities can lead to the development of instabilities in a weakly ionized plasma in crossed electric and magnetic fields (see, for example [18, 19]), which significantly reduce the efficiency of energy generation, or, in general, make the operation of the MHD generator impossible.

An example of a pulsed nanosecond discharge is a streamer characterized by fast-propagating ionization fronts with self-organized electric field enhancement at its tip. Streamer discharges have been much studied over many decades (see [20, 21] and references therein). However, little is known about the effect of magnetic field on streamer properties and the number of works devoted to the development of pulsed discharges in strong magnetic fields is small. In [22-24], the effect of a longitudinal magnetic field on the evolution of an electron avalanche and the time of breakdown formation was studied. Gas breakdown in weak longitudinal and transverse magnetic fields was investigated in [25]. In [26, 27], it was shown that a longitudinal magnetic field can suppress the development of the near-electrode instability in a non-self-sustained gas discharge. It was experimentally observed that the streamer channels can be deflected under a transverse magnetic field when propagating in a free gas [28] and along a dielectric surface [29, 30]. In all these studies, as a rule, the development of a discharge at high pressures and relatively weak magnetic fields was addressed, when the effect of a magnetic field was mainly reduced to a displacement of the discharge channel at long times. Only in [29] the propagation of a surface discharge at a relatively low (100 Torr) pressure and strong magnetic field ($B = 3$ T) was studied. However, in the geometry of [29], the magnetic field was perpendicular to the electric field, which led to simply bending of the surface streamer trajectories. Surface discharges at lower magnetic fields and higher pressures were also investigated in [30].

In [31], an attempt was made to develop an analytical model of a streamer in a longitudinal magnetic field, taking into account the change in the electron mobility across the magnetic field lines. In this study, the influence of the magnetic field on the average electron energy and electron impact ionization rate was not considered. Nevertheless, only taking into account the dependence of the electron mobility on the electric and magnetic field direction, the model led to the conclusion about a possible decrease in the radius of curvature of the streamer head and an increase in its propagation velocity when an external magnetic field is applied [31].

In the present work, a numerical characterization of nanosecond pulsed discharges has been conducted in a strong magnetic field environment. Streamer discharge development and plasma generation in pure $CO_2$ was analyzed when magnetic field was directed along the axis of the discharge cell. Numerical simulations were based on a two-dimensional fluid model. It was shown that a strong magnetic field affects dramatically the plasma formation. The nanosecond streamer diameter decreased significantly, whereas the plasma density increased with increasing the magnetic field amplitude. The mechanisms of the magnetic field interaction with the pulsed discharge development were discussed.

**Numerical model of nanosecond discharge development**

To study the effect of a longitudinal magnetic field on streamer properties, we considered a solitary streamer developing in the 14 cm plane-to-plane discharge gap. A streamer was initiated near the high-voltage electrode and propagated to the ground plane electrode. The high-voltage electrode was a plate at $Z = 1$ cm with a semi-ellipsoidal needle (a major semi-axis of 0.8 cm and a minor semi-axis of 0.08 cm) protruded from the center of the plate. The computational region was $15 \times 15$ cm$^2$ (Figure 1).

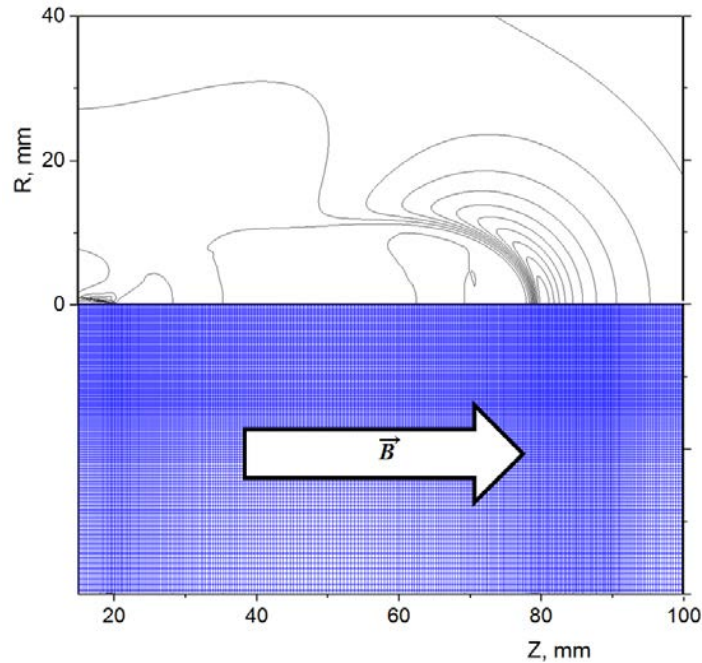

Figure 1. Discharge gap geometry and adaptive computational mesh (every 10$^{th}$ cell is shown). The tip of the high-voltage electrode is located at $Z = 20$ mm and the grounded plane electrode is located at $Z = 150$ mm. The top part of the figure shows the contours of the electric field of the positive streamer 40 ns after the start of the high voltage pulse.
$P = 50$ Torr, $T = 293$ K, $U = +20$ kV, $CO_2$.

The minimal grid size was 1.5 μm, whereas the minimal time step was $5\times10^{-14}$ s. The adaptive grid changed with streamer development in the gap. The cell size increase was limited by 10% to avoid the numerical error increase on a nonuniform grid. Calculations were carried out for different magnetic field values at fixed gas pressure $P = 50$ Torr, room gas temperature and fixed nanosecond pulse voltage $U = 20$ kV. We assumed that the applied voltage increased linearly to 20 kV for 1 ns and remained constant for $t > 1$ ns.

Streamer initiation and propagation was simulated on the basis of an axially symmetric 2D fluid model [32-37]. The system of equations under study consisted of the transport equations for the densities of charged particles (electrons and positive and negative ions) and Poisson's equation for the electric field. The transport and kinetics of charged species were considered in the local approximation, whereas the non-local approach was utilized to describe photoionization generating seed electrons in front of the streamer head by ionizing radiation emitted from the head and the streamer channel:

$$\frac{\partial n_e}{\partial t} + \text{div}(\vec{u_e} \cdot n_e) = S_{ion} + S_{photo} - S_{att} - S_{rec}^{ei} \quad (1)$$

$$\frac{\partial n_p}{\partial t} = S_{ion} + S_{photo} - S_{rec}^{ei} - S_{rec}^{ii} \quad (2)$$

$$\frac{\partial n_n}{\partial t} = S_{att} - S_{rec}^{ei} \quad (3)$$

$$\Delta\varphi = -\frac{e}{\varepsilon_0}(n_p - n_e - n_n) \quad (4)$$

where $S_{ion}$ is the ionization rate, $S_{photo}$ is the rate of photoionization, $S_{att}$ is the rate of electron attachment, and $S_{rec}^{ei}$ and $S_{rec}^{ii}$ are the rates of electron-ion and ion-ion recombination, respectively. Ionization of gas molecules by electron impact, electron attachment to molecules, and electron-ion and ion-ion recombination were taken into account. The photoionization term $S_{photo}$ has been taken according to the data presented in [38].

**Magnetic field influence on EEDF**

The electron impact ionization rate coefficient, the electron drift velocity $v_e$, and the effective electron temperature $T_e$ were calculated by solving the Boltzmann kinetic equation in the two-term approximation using the BOLSIG+ code [39, 40] and the self-consistent set of the electron cross sections in $CO_2$ [41]. In this approximation, the effect of magnetic field $\vec{B}$ on the

Boltzmann equation is reduced to the replacement of the electric field by the effective electric field [42]

$$E_{eff} = \sqrt{E_{\|}^2 + \frac{E_{\perp}^2}{[\beta_e(\varepsilon)]^2 + 1}}, \quad (5)$$

where $E_{\|}$ and $E_{\perp}$ are the longitudinal and transverse components of the electric field $\vec{E}$ with respect to the magnetic field $\vec{B}$, respectively, $\beta_e(\varepsilon) = \omega_e/\nu_m(\varepsilon)$ is the Hall parameter for electrons with energy $\varepsilon$, $\nu_m(\varepsilon) = n\, u(\varepsilon)\, Q_m(\varepsilon)$ is the momentum transfer frequency for electrons with energy $\varepsilon$, $u(\varepsilon)$ is the electron velocity, $Q_m(\varepsilon)$ is the electron momentum cross section and.

From (5), the effect of magnetic field on the electron energy distribution function (EEDF) is absent when this field is parallel to the electric field ($E_{\perp} = 0$). This effect is most profound when the vector $\vec{B}$ is perpendicular to the vector $\vec{E}$ and $\beta_e \gg 1$. In this case, the application of magnetic field inhibits drastically electron heating by the electric field and the effective electric field $E_{eff}$ decreases.

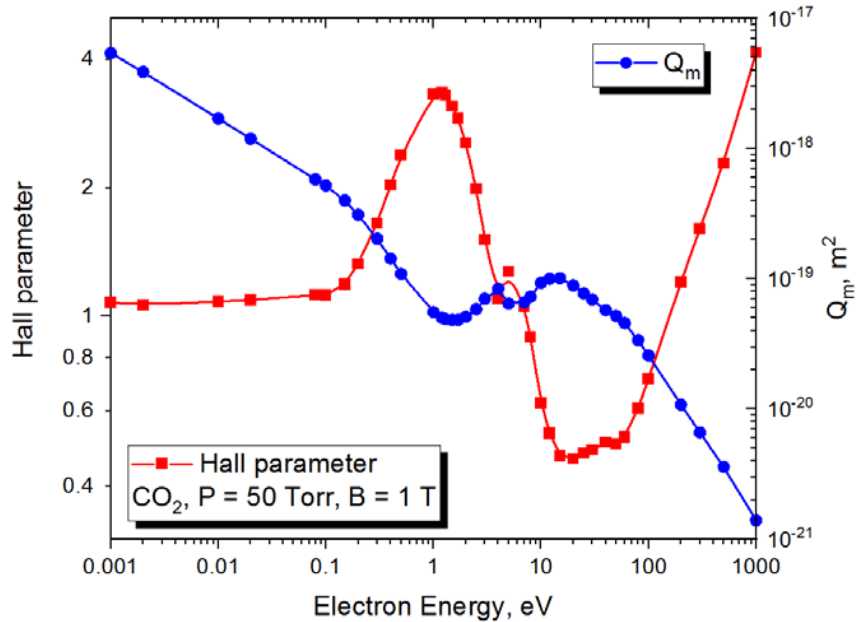

Figure 2. Electron momentum cross section and Hall parameter versus electron energy for $CO_2$ at $P = 50$ Torr and $B = 1$T.

Magnetic field affects equally all electrons when $Q_m(\varepsilon) \sim \varepsilon^{-0.5}$. Under such a condition the frequency $\nu_m(\varepsilon)$ and the Hall parameter are independent of $\varepsilon$. Figure 2 shows the electron momentum transfer cross section $Q_m$ for $CO_2$ and the corresponding Hall parameter versus the electron energy for $P = 50$ Torr, room gas temperature and $B = 1$ T. Because of a complicated energy dependence of $Q_m(\varepsilon)$ in $CO_2$, the effect of magnetic field is most profound for the electrons with energies between 0.1 and 4 eV and is the smallest one for the electrons with energies in the range 4 – 150 eV. As a result, the shape of the EEDF changes under the action of magnetic field. Figure 3 shows the calculated EEDF in $CO_2$ for 300 Td, 50 Torr, room gas temperature and various values of magnetic field when it is perpendicular to the electric field. From this figure, the slopes of the curves at sufficiently high $B$ are more gradual for $\varepsilon > 4$ eV in comparison with the slopes for lower electron energies. The shape of EEDF influences the electron transport and ionization rate coefficients which control a streamer development in crossed electric and magnetic fields.

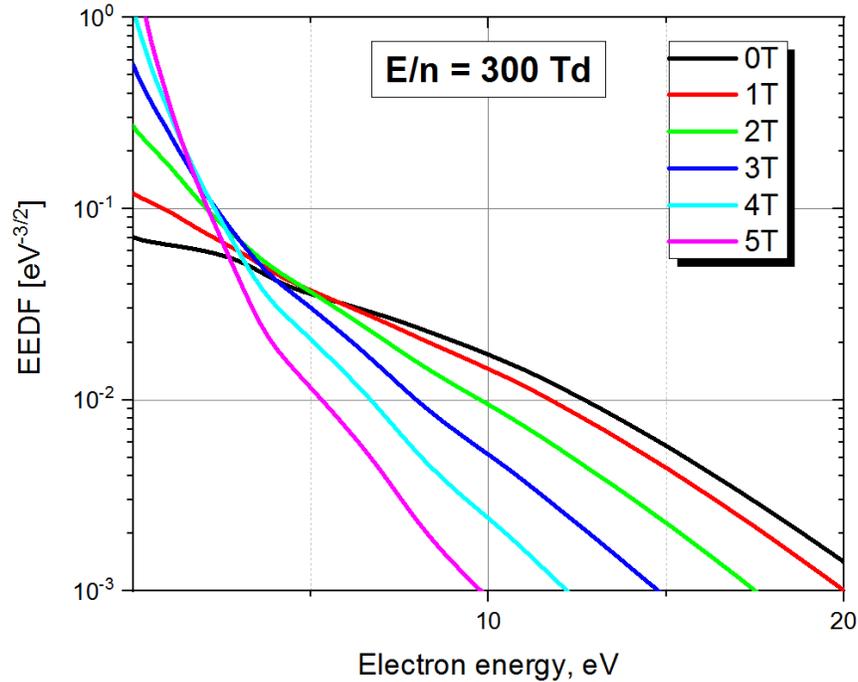

Figure 3. Electron energy distribution functions in $CO_2$ for 300 Td, $P = 50$ Torr and various values of magnetic field. The electric and magnetic fields are perpendicular to each other.

In a general case, the electron properties depend not only on the values of electric and magnetic fields, but on the angle $\alpha$ between these vectors as well. To take into account all possible combinations of $\vec{B}$ and $\vec{E}$, we generated a look-up table over $B$, $E$, and $\cos(\alpha)$ values, and then constructed analytical functions to interpolate the rate and mobility coefficients in the entire range of the parameters *B*, *E*, and cos(α). This approach dramatically reduced a computation time.

Figure 4 shows the result of the combined effect of electric and magnetic fields on the electron ensemble characteristics. EEDF and electron properties are not affected by magnetic field if the magnetic field vector $\vec{B}$ is collinear to the electric field vector $\vec{E}$ (cos(α) = 1). When the electric field vector is perpendicular to the magnetic field vector (cos(α) = 0), the influence of the magnetic field becomes very strong and increases with the decrease of the electric field value.

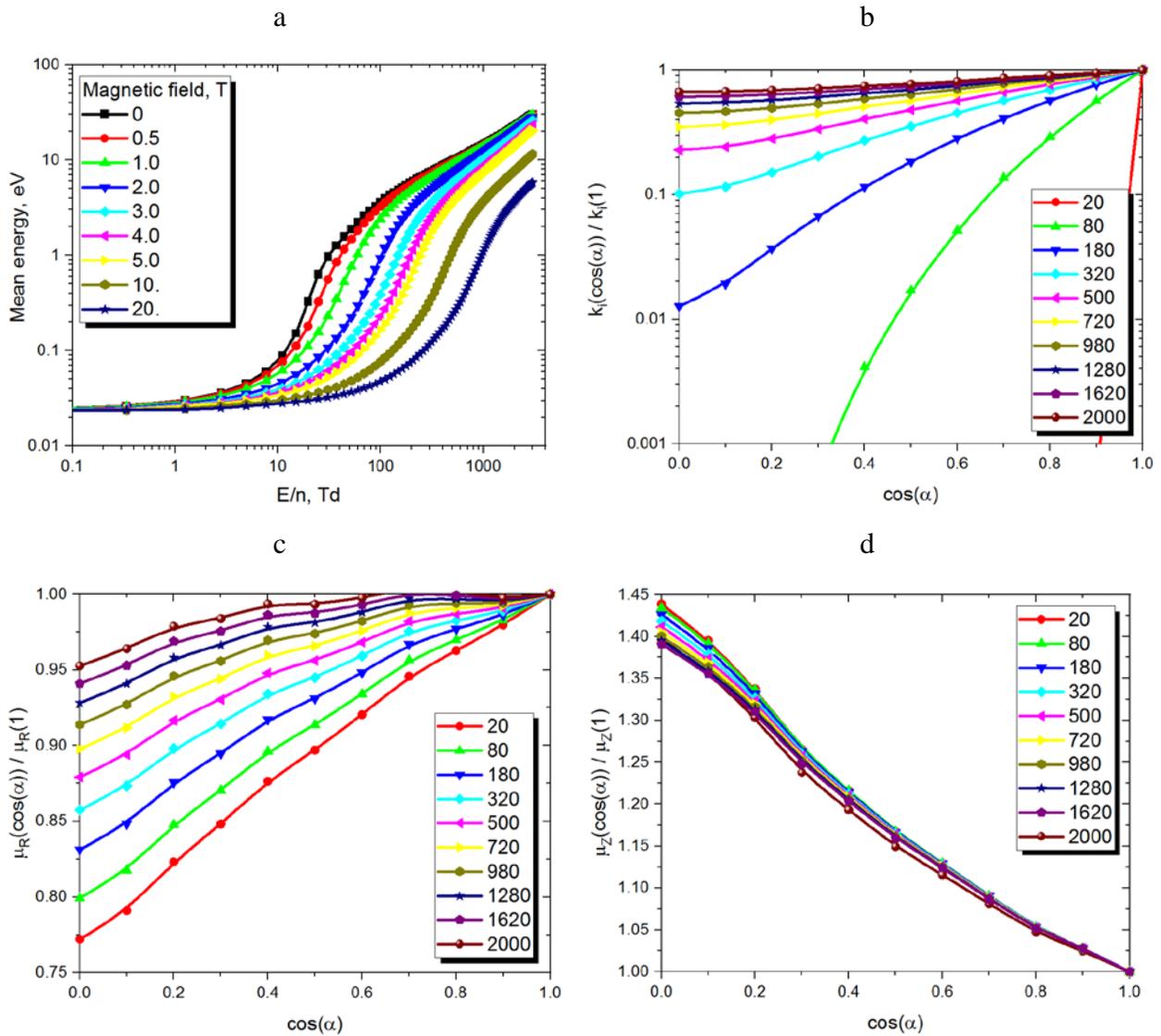

Figure 4. (a) Mean electron energy versus *E/n* for cos(α) = 0 and different *B*. (b) – (d) Ionization rate coefficient and electron mobility along the magnetic field vector ($\mu_Z$) and across the magnetic field vector ($\mu_R$) versus cos(α) for *B* = 3 T and different *E/n* (scales in Td).

Figure 4(a) demonstrates the effect of an external magnetic field on the average electron energy in the case of perpendicular electric and magnetic fields ($\cos(\alpha) = 0$). In the absence of magnetic field, the average electron energy $\langle\varepsilon\rangle$ begins to grow rapidly at $E/n > 10$ Td and reaches 1 eV already at $E/n \sim 31$ Td. An increase in the magnetic field leads to a decrease in $\langle\varepsilon\rangle$. The electric field required to obtain $\langle\varepsilon\rangle = 1$ eV is $E/n \sim 58$ Td for $B = 1$T, 147 Td for $B = 3$ T and 960 Td for $B = 20$ T. Such a strong dependence of the average electron energy on the magnitude of the perpendicular magnetic field obviously leads to a strong decrease in the rate of gas ionization by electron impact in an ionization wave propagating across the magnetic field lines. At the same time, the same magnetic field directed along the electric field vector does not cause any change in the ionization rate (Figure 4(b)). It can be seen that the decrease in the ionization rate in a radial ionization wave (propagating across the magnetic field lines) compared to a longitudinal wave (propagating along the magnetic field lines) is two orders of magnitude at $E/n = 180$ Td, and one order of magnitude at $E/n = 320$ Td. Even in a very strong electric field ($E/n \sim 1000$ Td), a twofold decrease in the ionization rate in the radial wave compared to the longitudinal wave is observed for a relatively weak magnetic field of $B = 3$ T. The effect is enhanced with the magnetic field increase and weakens with increasing $E/n$ (Figure 4(b)).

Another consequence of the appearance of a strong magnetic field directed along the streamer axis is a sharp decrease in the electron mobility in the radial (across the magnetic field lines) direction (Figure 4(c)). As in the case of the ionization coefficient, the greatest decrease in the mobility is obtained in weak electric fields. At high $E/n$, the effect of the transverse magnetic field on the electron drift in the radial direction weakens. At the same time, the electron mobility in the direction of the magnetic field only slightly depends on the magnitude of the electric field and strongly depends on the direction of the magnetic field relative to the electric one (Figure 4(d)).

Such a strong influence of the magnetic field on the electron ensemble characteristics, of course, should affect the development of discharges in the gas. In the next section, we will consider this effect for a streamer discharge.

**Streamer propagation dynamics in strong external magnetic field**

We made a numerical modeling of the development of a nanosecond streamer discharge in a strong external magnetic field directed along the axis of the discharge gap.

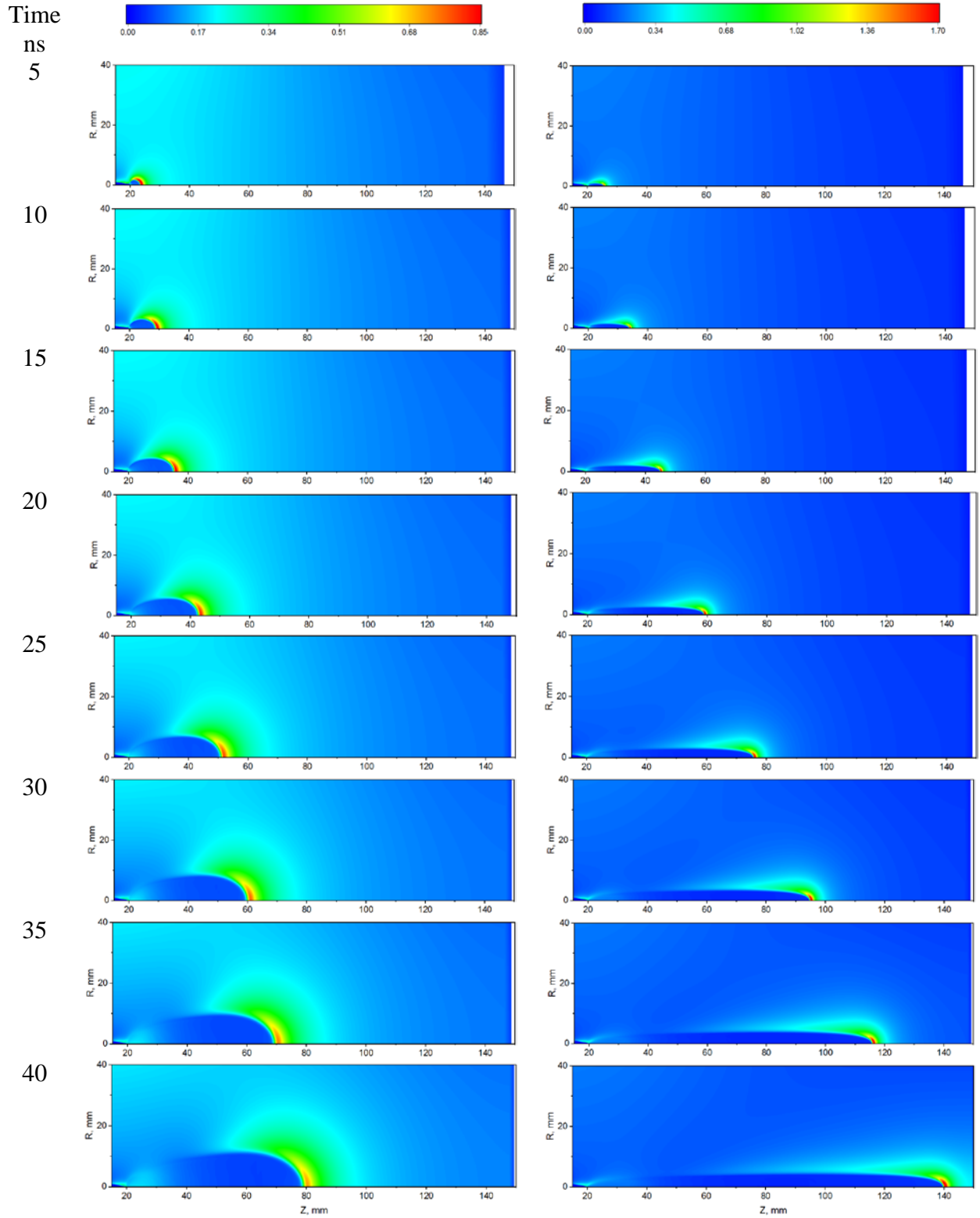

Figure 5. Spatial electric field distribution during the development of a positive streamer for $B = 0$ (left) and $B = 3$ T (right). Left scale: 0-0.85 MV/m. Right scale: 0-1.7 MV/m. $P = 50$ Torr, $CO_2$, $U = +20$ kV.

The calculations were carried out for $CO_2$ pressure of $P = 50$ Torr, room gas temperature and a voltage of $U = 20$ kV across the gap. The magnitude of the magnetic field was varied from 0 to 20T, which corresponds to the reduced gyrofrequency

$$\omega_e/n = \frac{eB}{m_e n} = 0 - 20 \times 10^{-13} \text{ [rad} \times \text{m}^3\text{/s]}.$$

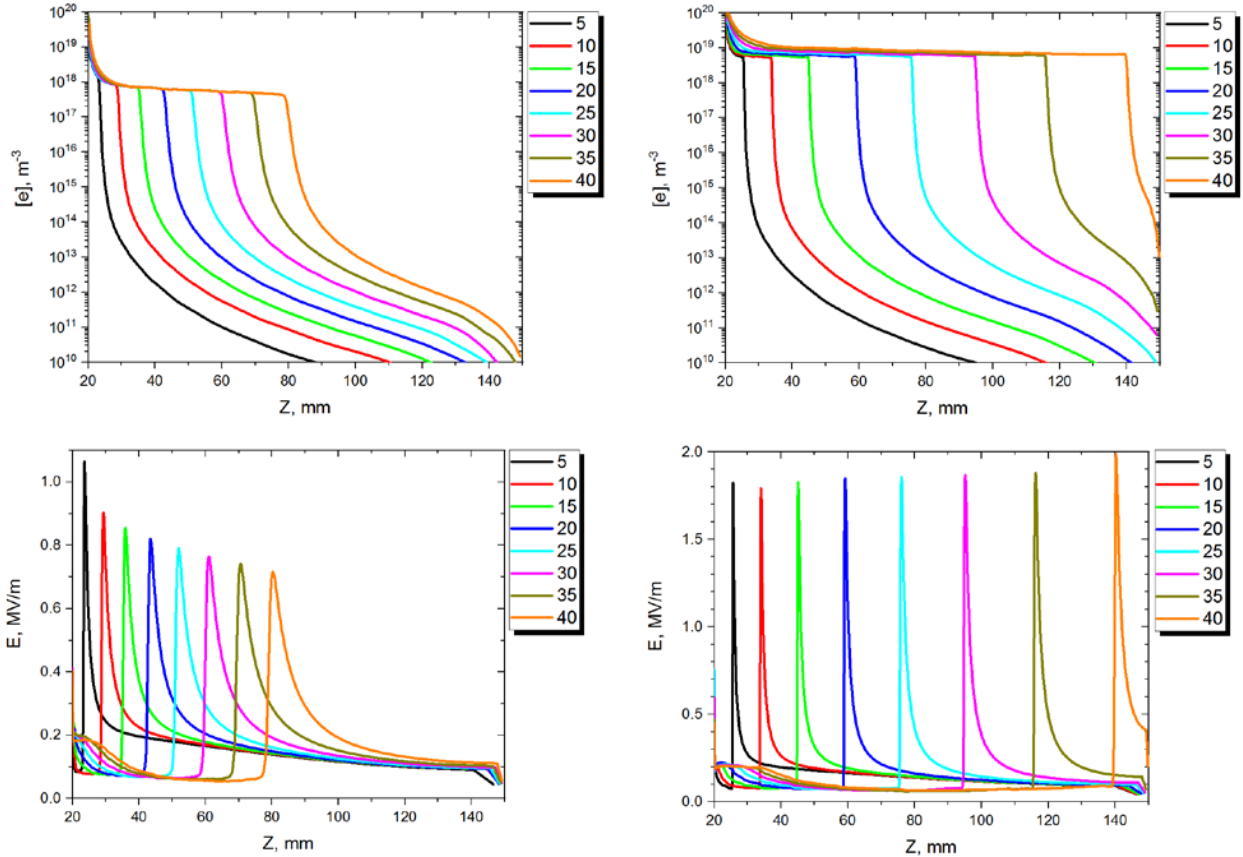

Figure 6. Temporal evolution of the axial profiles of the electron density (upper row) and electric field (bottom row) during positive streamer development for $B = 0$ (left) and $B = 3$ T (right). Calculations are for $P = 50$ Torr, $CO_2$, and $U = +20$ kV. The legend shows time from the discharge start in nanoseconds.

Figure 5 shows the dynamics of the spatial electric field distribution during the development of a positive streamer for $B = 0$ (left) and $B = 3$ T (right). From the figure, the radial ionization wave development is significantly suppressed in the presence of the magnetic field. As a result, the streamer diameter sharply decreases, whereas its propagation velocity grows. The electric field at the streamer head almost doubles. This behavior of the streamer discharge in the longitudinal magnetic field can be interpreted as a self-focusing effect, because the discharge modified the gas medium such that the streamer diameter gradually decreased during the discharge propagation.

Figure 6 shows the temporal evolution of the axial profiles of the electron density and electric field during the positive streamer development in a magnetic field and in its absence. The streamer developing in the strong magnetic field has an almost an order of magnitude higher electron density in the channel (upper row), a higher electric field on the head, and much slower attenuation as it moves through the gap. At the same time, the electric field in the streamer channel remains almost the same in both cases. The electric field in the streamer channel is $E = 0.56$ kV/cm at $B = 0$ and $E = 0.62$ kV/cm at $B = 3$ T. An increase in the electron density in the channel and an increase in the plasma conductivity with the streamer development in the magnetic field leads to a decrease in the electric field in the channel. However, this electric field decrease is compensated by the electric field enhancement due to an increase in the streamer speed and a decrease in the channel diameter. As a result, the magnitude of the electric field in the streamer channel is almost the same with the magnetic field and without it (Figure 6).

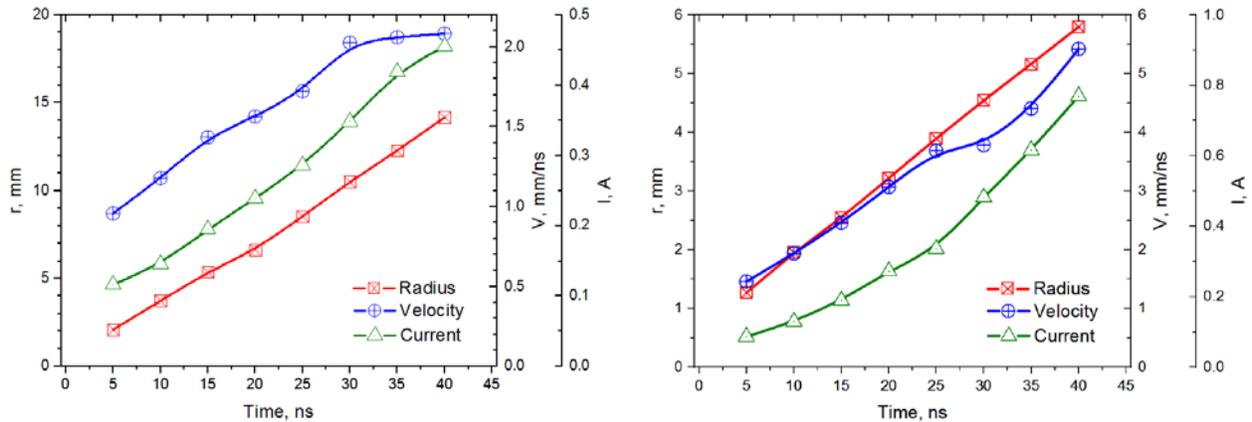

Figure 7. Temporal evolution of the channel radius, propagation velocity, and discharge current for a positive streamer propagating at $B = 0$ (left) and $B = 3$ T (right).
$P = 50$ Torr, $CO_2$, $U = +20$ kV.

Figure 7 compares the channel radius, propagation velocity, and discharge current for a positive streamer propagating at $B = 0$ and $B = 3$ T. The radius of the streamer channel propagating without magnetic field is 2.5 times that of the streamer channel in the magnetic field. The opposite relation is observed for the propagation velocity. The streamer in the magnetic field propagates 3 times faster than the streamer at $B = 0$. As a result, the streamer current in the magnetic field is double that without magnetic field (Figure 7).

| Time, ns | Hall parameter / Mean electron energy |
|---|---|
| 20 | 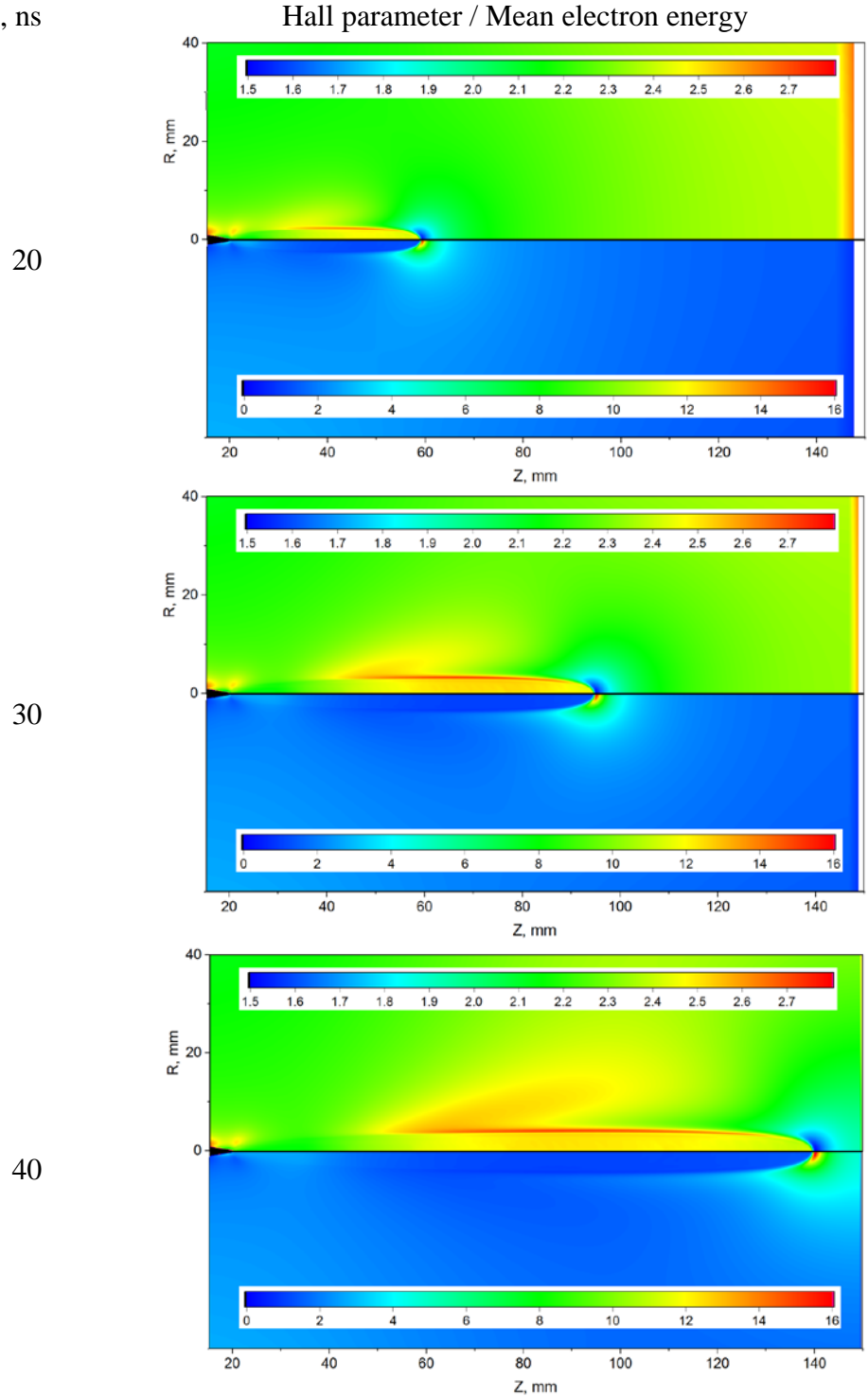 |
| 30 | |
| 40 | |

Figure 8. Temporal evolutions of the spatial distributions of the Hall parameter (upper part of the image) and electron mean energy (lower part of the image) for a positive streamer developing in a magnetic field of $B = 3$ T. Scale for Hall parameter: 1.5-2.8. Scale for $\langle\varepsilon\rangle$: 0-16 eV. $P = 50$ Torr, $CO_2$, $U = +20$ kV.

As mentioned above, an important parameter that determines the interaction of an electron ensemble with a magnetic field is the Hall parameter $\beta(\langle\varepsilon\rangle) = \omega_e/\nu_e$, which is the ratio of the gyrotron frequency of electrons in a magnetic field to the total frequency of electron transport collisions. As a rule, the effect of magnetic field becomes significant if the Hall parameter is higher than unity. Figure 8 shows the spatial distribution of the Hall parameter $\beta(\langle\varepsilon\rangle)$ and the average electron energy $\langle\varepsilon\rangle$ when the streamer is moving in the gap with a magnetic field of $B = 3$ T. From this figure, the Hall parameter is relatively small at the streamer head where the magnetic field vector is parallel to the vector of the electric field, while the average electron energy and the electron transport frequency are high. As a result, here, we have $\beta(\langle\varepsilon\rangle) = 1.5\text{-}1.7$ (Figure 8).

Different values of the parameters under consideration are obtained on the lateral surface of the streamer channel. Here, the vector of the local electric field is perpendicular to the vector of the magnetic field. The electron energy and the ionization rate sharply decrease (Figure 8) and the Hall parameter becomes twice as large as that at the leading ionization wave. Thus, the Hall parameter increases from 1.5 at the streamer head to 3 on the lateral surface of the channel. This increase leading to a decrease in the ionization rate, as well as a low transverse electron mobility, clearly shows the mechanism of suppression of the radial ionization wave development during the streamer discharge propagation along a strong magnetic field in the discharge gap.

In the first 20 nanoseconds after the start, the streamer traveled 40 mm; that is, its average speed was 2 mm/ns (Figure 8). In this case, the average electron energy was 16 eV in the head on the streamer axis and 3 eV at the radial ionization wave on the lateral surface of the channel. Over the next 10 ns, the streamer traveled another 35 mm. Here, the average streamer speed increased to 3.5 mm/ns due to a reduction in the radius of the curvature of the leading ionization wave front and an increase in the average field in the gap. The ratio of the Hall parameter on the streamer head and that on the streamer lateral surface remained practically the same as in the previous interval. A pronounced side lobe of high values of the Hall parameter appeared near $Z = 50$ mm, which was associated with the spatial structure of the electric field in the discharge gap and a local decrease in the average electron energy far from the high-voltage electrode and the streamer head. The same distribution of the discharge parameters is preserved for $t = 40$ ns (Figure 8). The streamer velocity in the gap increased to 4.5 mm/ns due to the electric field enhancement when approaching the grounded electrode. The Hall parameter remained high ($\beta(\langle\varepsilon\rangle) \sim 3$) on the lateral surface of the streamer and approximately 50% lower on its head. Thus, the interaction of electrons with a magnetic field leads to a significant decrease in the average electron energy in the radial ionization wave, which prevents the streamer channel from expanding.

**Influence of discharge polarity**

Figure 9, $B = 0$, demonstrates the initial phase of the positive streamer development near the tip of the high-voltage electrode in the absence of magnetic field. At a voltage rise rate of 20 kV/ns, the streamer does not have time to noticeably increase the radius until the instant when the voltage reaches its maximum value. The electric field at the leading ionization wave becomes high and practically independent of direction. In the absence of magnetic field, this means a high average electron energy and a high ionization rate in the ionization wave for all directions of its propagation. As a result, the channel rapidly expands and the electric field at the streamer head decreases to a typical value of $E \sim 1$ MV/m, which corresponds to the reduced electric field $E/n \sim 600$ Td at $P = 50$ Torr.

In the case of a longitudinal magnetic field in the discharge gap, the local electric field on the lateral streamer surface is perpendicular to the magnetic field. Here, the average electron energy and the electron impact ionization rate are significantly reduced due to the interaction of the electron ensemble with the magnetic field. Under such conditions, the radial ionization wave is significantly suppressed. Its speed decreases and the streamer radius remains much smaller compared to that in the absence of the magnetic field. A decrease in the radius of the streamer head leads to an increase in the electric field on it, and an increase in the velocity of the longitudinal ionization wave. As a result, the streamer becomes thinner with an increase in the longitudinal magnetic field in the gap, and its speed increases with $B$ (Figure 9).

Figure 10 shows the development of a negative streamer at different magnetic fields in the gap. The negative streamer propagates faster than the positive streamer because of the low efficiency of photoionization, which controls the positive streamer propagation, at low gas pressure. Therefore, Figure 10 presents the spatial distributions of the electric field and electron density at $t = 4.2$ ns after the start of the discharge rather than at $t = 7.0$ ns, as in the case of the positive streamer.

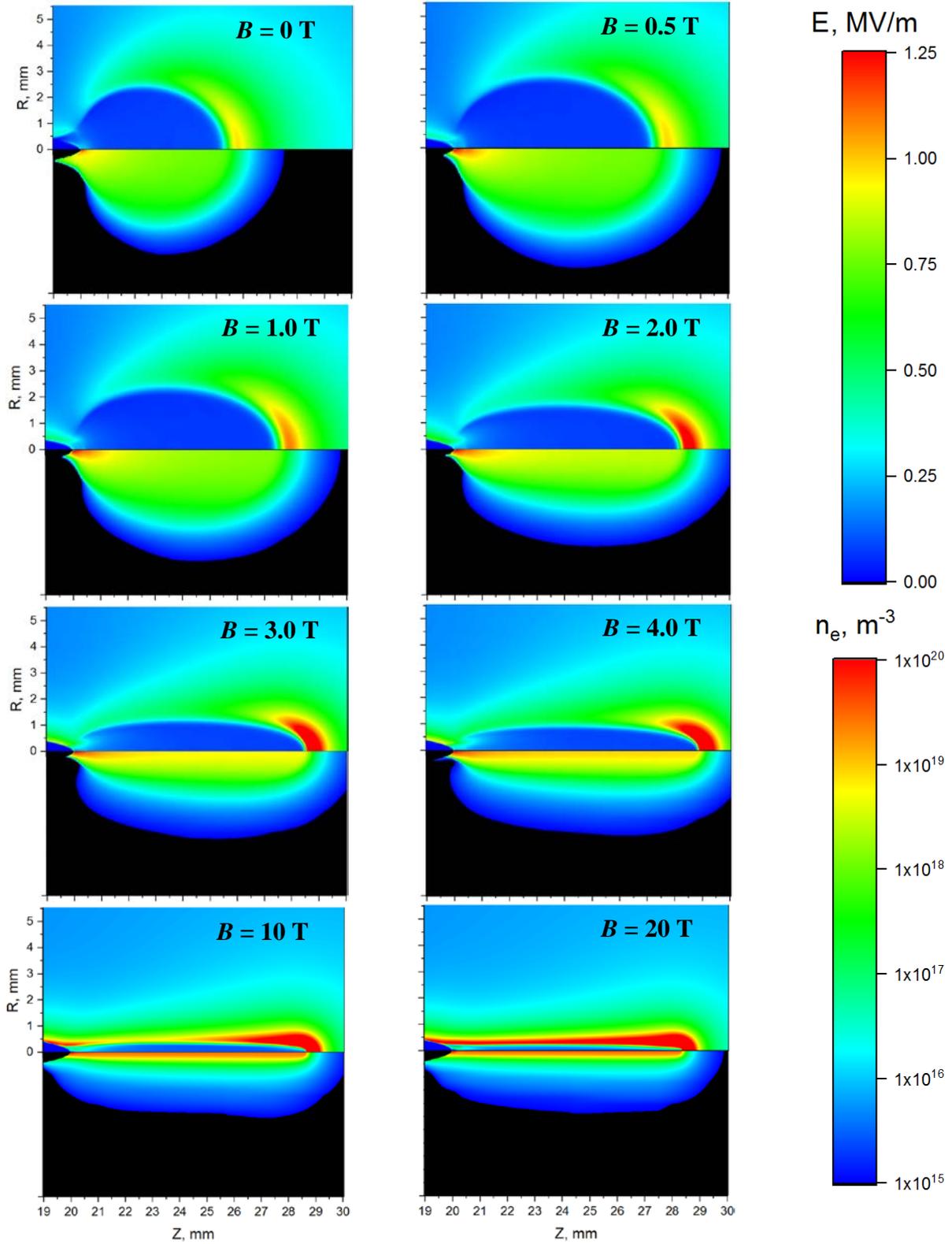

Figure 9. Spatial distributions of the electric field (upper part of the image) and electron density (lower part of the image) for the positive streamer propagation in different magnetic fields. $P = 50$ Torr, $CO_2$, $t = 7.0$ ns after high-voltage pulse start, $U = +20$ kV.

The general behavior of the negative streamer is similar to that of the positive streamer. When the magnetic field exceeds 1 T, the characteristics of the negative streamer change due to the effect of the magnetic field on the EEDF and integrated electron coefficients.

As in the case of the positive streamer, when the magnetic field increases in the gap, the radius of the streamer channel sharply decreases, whereas the electric field on the streamer head, the propagation velocity and the electron density in the channel increase (Figure 10).

However, there is an important difference between the propagation of negative streamers and positive ones. Strong negative streamers, as well as strong positive streamers, propagate due to photoionization of the gas ahead of the wave front and electron impact ionization in the high electric field at the streamer head [43]. In addition, negative streamers have the mechanism of propagation in the "weak" regime due to the forward electron drift, which is absent in the case of positive polarity [43]. This difference in the propagation mechanisms becomes especially important when the propagation distance of the photoionizing radiation becomes much larger than the characteristic radius of the streamer channel [43, 44]. This leads to a decrease in the efficiency of photoionization and significantly slows down the propagation of positive streamers. This effect should be especially pronounced at the highest values of the magnetic field, when the head radius of the propagating streamer is minimal.

Comparison of the streamer velocities calculated for positive and negative polarities of the high-voltage pulse clearly shows these different trends. The velocity of the negative streamer continuously increases as the magnetic field in the gap increases (Figure 10). The positive streamer sharply accelerates with an increase in the magnetic field for $B < 4$ T and slows down with increasing $B$ at $B > 4$ T. This deceleration is explained by a decrease in the efficiency of photoionization at small radii of the positive streamer head [43, 44].

The difference between the behavior of the positive streamer and negative one is more pronounced in Figure 11, which shows the axial profiled of the electron density and electric field for different discharge polarities and magnetic fields. The negative and positive streamers behave almost identically when the magnetic field increases from 0 to 4 T. Here, the streamer speed, the electric field on the head, and the electron density increase with $B$. A further increase in the magnetic field leads to opposite results for the streamers of different polarities. The positive streamer slows down with further increase in $B$.

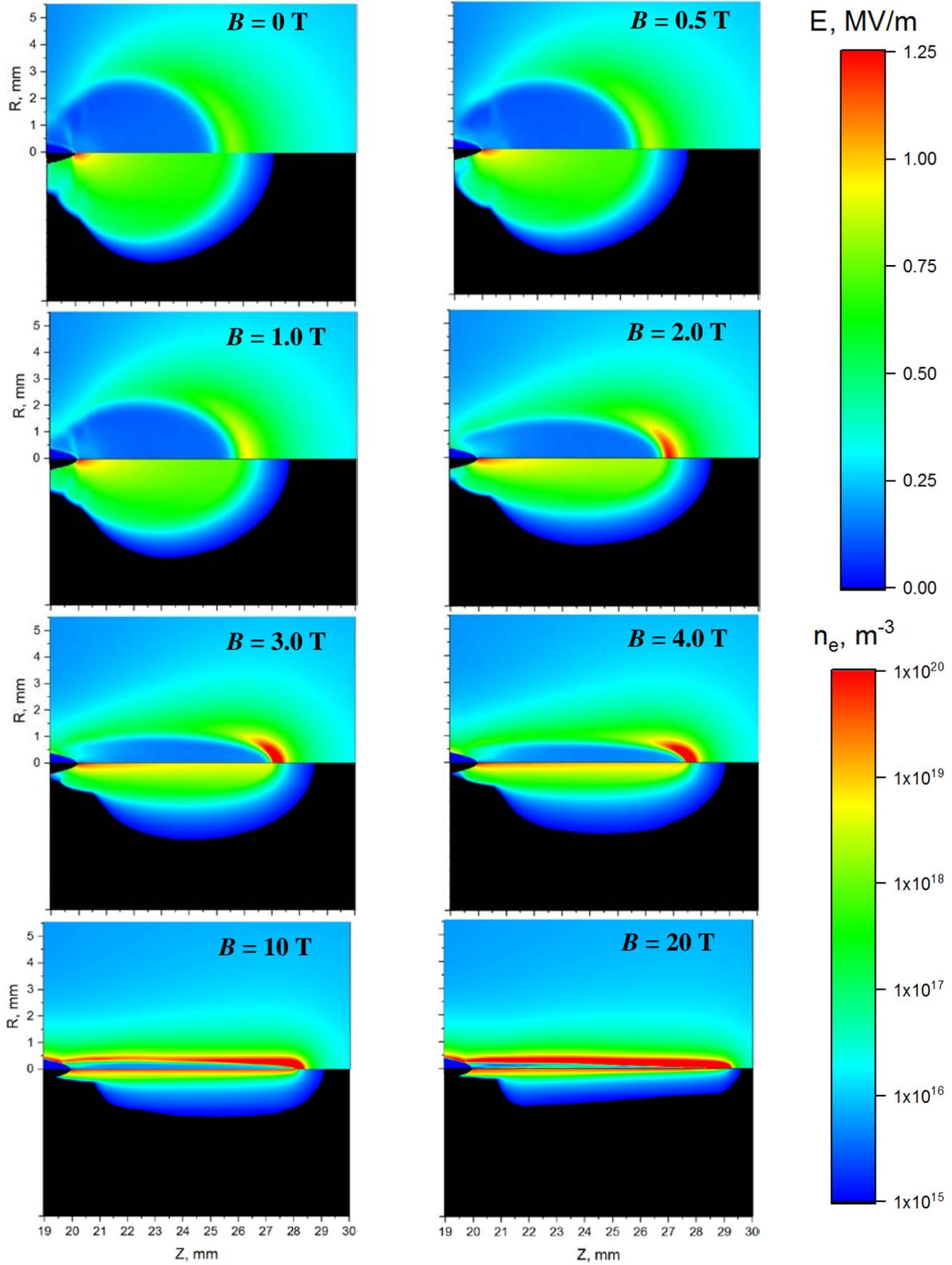

Figure 10. Spatial distributions of the electric field (upper part of the image) and electron density (lower part of the image) for the negative streamer propagation in different magnetic fields. $P = 50$ Torr, $CO_2$, $t = 4.2$ ns after high-voltage pulse start, $U = -20$ kV.

On the contrary, the negative streamer accelerates sharply and the electron density in the streamer channel increases with further increase of the magnetic fields. This is due to the positive feedback arising from the suppression of the radial ionization wave. In the strong magnetic field, the electrons can drift only forward along the axis of the discharge.

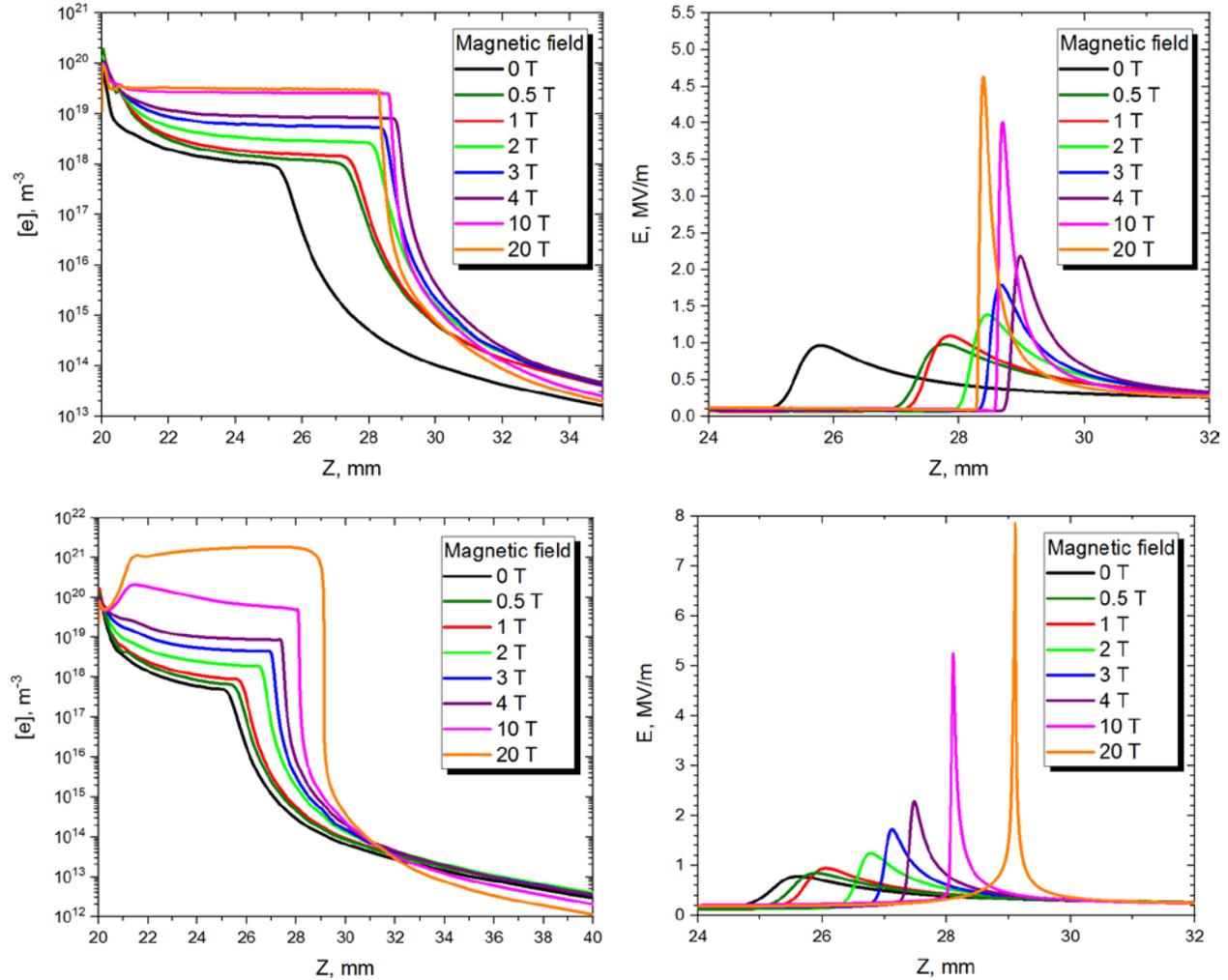

Figure 11. Axial profiles of the electron density and electric field for different discharge polarities and magnetic fields. Top: positive polarity, $t = 7.0$ ns. Bottom: negative polarity, $t = 4.2$ ns. $P = 50$ Torr, $CO_2$, $U = 20$ kV.

Such a displacement of the negative charge along the axis leads to an increase in the local axial electric field (Figure 11, bottom row) and favors the electron impact ionization on the discharge axis where the magnetic field is parallel to the electric field and does not reduce the ionization rate. The strong ionization leads to an increase in the electron density and an increase in the rate of charge transfer along the axis. Self-focusing of the streamer is observed, which causes a decrease in the effective radius of the streamer head and an increase in the local electric field in its vicinity.

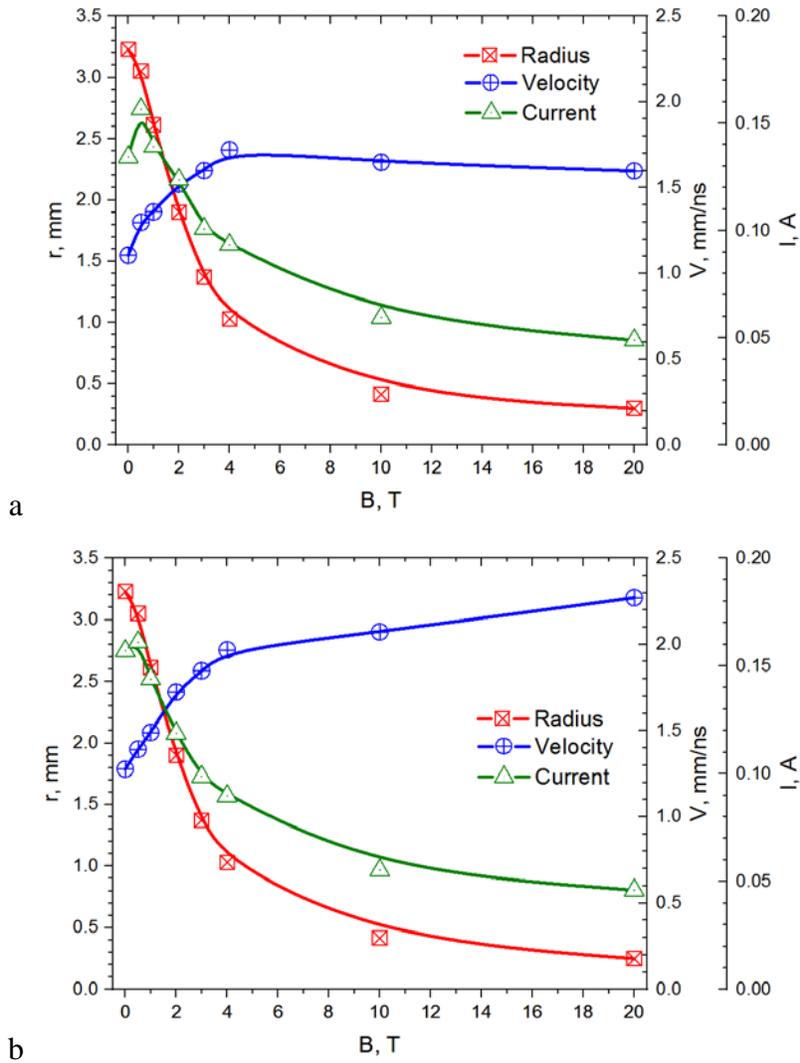

Figure 12. Streamer velocity, discharge current and channel radius versus magnetic field for different voltage polarities. A: positive polarity, $t = 7.0$ ns. B: negative polarity, $t = 4.2$ ns. $P = 50$ Torr, $CO_2$, $U = 20$ kV.

For the positive streamer, such a self-focusing is limited by the mean free path of UV photons and nonlocal gas preionization in front of the streamer. Thus, the direction of the electron drift in the streamer head determines the different behavior of the positive and negative streamers at the high magnetic fields (Figure 11). This conclusion has been made in [43, 44] for the streamer propagation in the discharge gap without magnetic field. The present results extend the conclusion made in [43, 44] to the case of strong external magnetic fields.

Figure 12 summarizes the data on the influence of the magnetic field on the radius of the streamer channel, discharge current and the velocity of streamer propagation. The magnetic field increase from 0 to 4 T leads to a decrease in the streamer radius almost by an order of magnitude. Here, the streamer propagation velocity increases monotonously. As the magnetic field increases from 0 to 4 T, the ionization wave propagation velocity almost doubles due to a decrease in the radius of the conducting channel and an increase in the local field at the streamer head (Figure 10). Figure 12 shows that the radial ionization wave weakens and the streamer radius decreases with increasing magnetic field because of the suppression of the radial electrons drift and significant decrease of the mean electron energy on the lateral streamer surface.

**Critical magnetic field**

It may be concluded that the streamer discharge sharply changes its characteristics with increasing the magnetic field from 0 to 4 T. With further increase in the magnetic field, the discharge parameters change much more smoothly. Thus, the reduced electron gyrofrequency $\omega_e/n \sim 10^{-13}$ [rad×m$^3$/s] corresponding to $B = 4$ T at 50 Torr is an important threshold above which the radial expansion of a streamer discharge in $CO_2$ is significantly suppressed. It should be noted that the reduced electron gyrofrequency is not a universal criterion for determining the threshold magnetic field, which allows one to significantly control the development of a pulsed high-voltage discharge. This value could change significantly due to different electron momentum transfer cross sections $Q_m(\varepsilon)$ in gases and different dependencies of the average electron energy on $E/n$. The most appropriate criterion should be based on the Hall parameter, which is the ratio of the electron gyrofrequency to the electron momentum transfer frequency. As noted above, the effect of the magnetic field is significant when the Hall parameter exceeds unity. However, to calculate the value of the Hall parameter, it is required to know the distribution of the electric field in the discharge gap (Figures 3 and 8). Generally speaking, a complete calculation of the discharge development in the magnetic field should be made. At the same time, the reduced electron gyrofrequency $\omega_e/n$ depends only on the initial parameters of the problem and could be a convenient estimate of the possible influence of external magnetic field on the discharge development, by analogy with the reduced electric field $E/n$ in the gap.

By extrapolating the values obtained in this work to normal conditions (atmospheric pressure and room gas temperature), it can be estimated that a longitudinal magnetic field significantly affects the development of a high-voltage discharge only at $B > 15$ T and that the magnetic field controls the development of such a discharge at high pressure for $B > 60$ T. These considerations limit the area of application of such methods to relatively low gas densities or ultra-small discharge gaps with strong magnetic fields. At the same time, for pressures up to 100 Torr, even relatively low magnetic fields in the gap can dramatically change the characteristics of a pulsed discharge development, causing strong discharge self-focusing.

**Conclusions**

Thus, based on the results of numerical simulation of the development of a streamer discharge in $CO_2$ in a gap with an external longitudinal magnetic field, the possibility of self-focusing of such discharges is demonstrated. The self-focusing is caused by a sharp slowdown in the speed of the radial ionization wave due to a change in the EEDF, a decrease in the average electron energy, electron mobility, and the rate of electron impact ionization in the crossed electric and magnetic fields as compared with the case of the discharge development without magnetic field. Simultaneously with the deceleration of the radial ionization wave, the ionization wave accelerates along the axis of the discharge gap due to a decrease in the radius of the streamer head and an increase in the electric field on it. Since the electric and magnetic fields are parallel to each other on the axis of symmetry, for the longitudinal wave there is no decrease in the average electron energy and ionization rate with an increase in the magnetic field value.

In a weakly ionized plasma, a transverse magnetic field slows down the heating of electrons in an external electric field. The effect of magnetic field on the electron properties is determined by the Hall parameter and turns out to be different for electrons belonging to different parts of their energy distribution function. In particular, in weakly ionized $CO_2$ plasma, the effect of magnetic field is most profound for electrons with energies in the 0.1–4 eV range and is less important for electrons with higher energies.

The self-focusing effect of a streamer discharge in an external longitudinal magnetic field is observed for both polarities of the discharge. At the same time, the self-focusing of the positive streamer is limited by a decrease in the photoionization efficiency as the streamer radius decreases to values less than the propagation distance of ionizing radiation in the gas. This limitation is absent for the negative streamer.

The estimates of the critical magnetic field obtained in this work lead to the conclusion that the effective interaction of the external magnetic field with the ionization wave occurs when the Hall parameter exceeds unity. For a qualitative assessment of this effect, it is proposed to consider the critical value of the reduced electron gyrofrequency $\omega_e/n \sim 10^{-13}$ rad×m$^3$/s (in $CO_2$), which may be somewhat different for other gases.

**Acknowledgements**

This work was supported by DOE grant DE-SC0021330, Texas A&M University/DOE–NETL grant DE-FE0026825 and by DOE grant DE-FE0026825.